\documentclass[10pt]{iopart}
\usepackage{setstack}
\usepackage{iopams}
\usepackage{graphicx}

\begin{document}

\title[]{Non-equilibrium relaxation in a stochastic lattice Lotka-Volterra model}

\author{Sheng Chen and Uwe C. T\"auber}
\address{Department of Physics (MC 0435), Robeson Hall, 850 West Campus Drive, 
		 Virginia Tech, Blacksburg, Virginia 24061, USA}
\ead{csheng@vt.edu, tauber@vt.edu}

\date{\today}

\begin{abstract}
We employ Monte Carlo simulations to study a stochastic Lotka-Volterra model on 
a two-dimensional square lattice with periodic boundary conditions. 
If the (local) prey carrying capacity is finite, there exists an extinction 
threshold for the predator population that separates a stable active two-species
coexistence phase from an inactive state wherein only prey survive.
Holding all other rates fixed, we investigate the non-equilibrium relaxation of 
the predator density in the vicinity of the critical predation rate.
As expected, we observe critical slowing-down, i.e., a power law dependence of
the relaxation time on the predation rate, and algebraic decay of the predator
density at the extinction critical point. 
The numerically determined critical exponents are in accord with the established
values of the directed percolation universality class. 
Following a sudden predation rate change to its critical value, one finds 
critical aging for the predator density autocorrelation function that is also
governed by universal scaling exponents. 
This aging scaling signature of the active-to-absorbing state phase transition 
emerges at significantly earlier times than the stationary critical power laws, 
and could thus serve as an advanced indicator of the (predator) population's 
proximity to its extinction threshold. 
\end{abstract}


\vspace{2pc}
\qquad\qquad 	critical dynamics, aging scaling, early warning signals.

\submitto{\PB -- \today}


\maketitle

\ioptwocol

\section{Introduction}

There is growing interest in quantitatively understanding biodiversity in 
ecology \cite{r1973, j1974} and population dynamics \cite{jk1998, j2002, n2004}.
The motivations for this very active research field range from seeking a 
fundamental and comprehensive understanding of noise-induced pattern formation 
and phase transitions in far-from-equilibrium systems to potential practical
applications in protecting endangered species in threatened ecosystems. 

Unfortunately, the full complexity of interacting species in coupled ecosystems 
in nature cannot yet be reliably modeled with the required faithful incorporation 
of demographic fluctuations and internal stochasticity induced by the involved 
reproduction and predation reactions.
One therefore typically resorts to detailed investigations of idealized,
simplified models that however are intended to capture the important system
ingredients and ensuing characteristic properties.
The Lotka--Volterra predator-prey model \cite{a1920, v1926} has served as such a 
simple but intriguing and powerful paradigm to study the emerging coexistence of 
just two species, predators and prey, as a first step to grasp the initially 
counter-intuitive appearance of biodiversity among competing species. 
In the model's original formulation, the authors just analyzed the associated 
coupled mean-field rate equations, whose solution remarkably entails a stable 
active coexistence state in the form of a neutral cycle: the densities of both 
populations hence display periodic non-linear oscillations. 

\begin{figure*}[t]
\centering
\includegraphics[width=0.66\columnwidth,height=0.66\columnwidth]{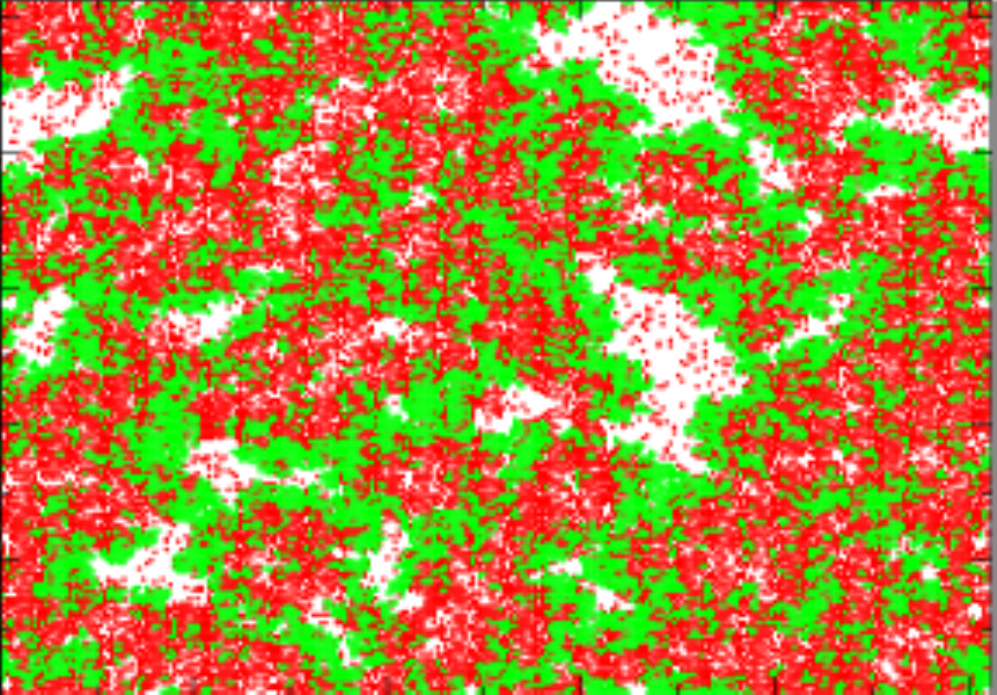} \
\includegraphics[width=0.66\columnwidth,height=0.66\columnwidth]{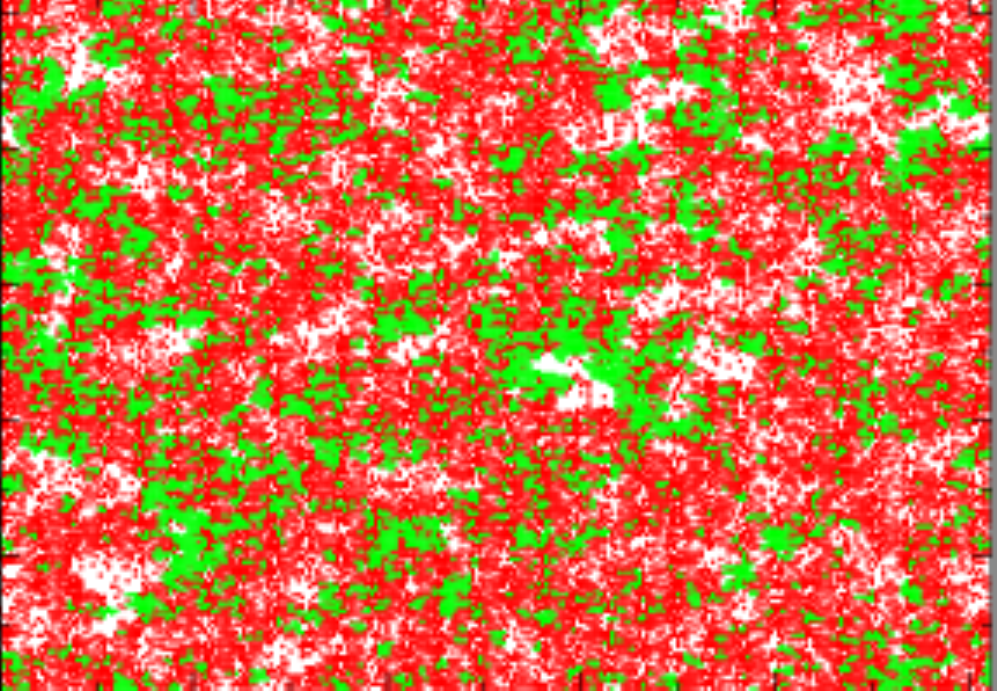} \
\includegraphics[width=0.66\columnwidth,height=0.66\columnwidth]{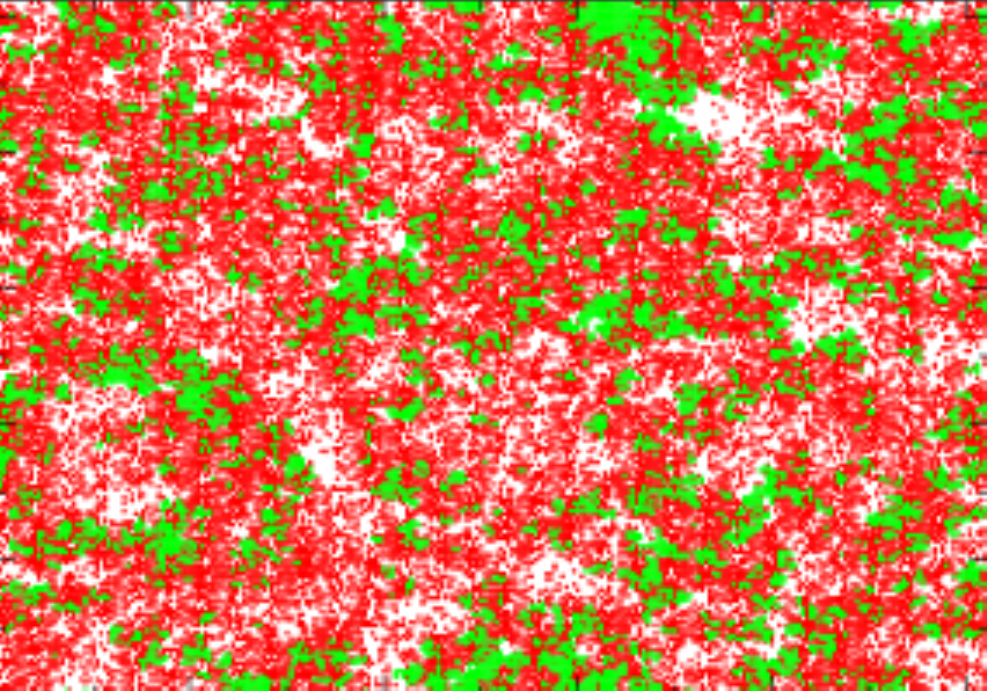}
\caption{Snapshots of the spatial particle distribution for a single Monte Carlo 
		simulation run of a stochastic Lotka--Volterra model on a $256 \times 256$
		square lattice with periodic boundary conditions, successively from left
		to right at $t = 500$~Monte Carlo steps (MCS), $t = 1000$~MCS, and 
		$t = 2000$~MCS, with reaction probabilities (see text) $\mu = 0.025$, 
		$\lambda = 0.25$, and $\sigma = 1.0$; only at most one particle per 
		lattice site is allowed: sites occupied by predators are indicated in red,
		prey in green, while empty sites are shown white.}
\label{fige1}
\end{figure*}

Yet this classical deterministic Lotka--Volterra model has been aptly criticized 
for its non-realistic feature of the oscillations being fully determined by the
system's initial state, and for the lack of robustness of the marginally stable 
neutral cycle against model perturbations \cite{j2002}. 
The importance of stochasticity as well as spatio-temporal correlations, both
entirely neglected in the mean-field approximation, was subsequently recognized 
in a series of numerical simulation studies of several stochastic spatially 
extended lattice Lotka-Volterra model variants \cite{hnak1992}--\cite{mmu2007}.
Even in the absence of spatial degrees of freedom, stochastic Lotka--Volterra 
models display long-lived but ultimately decaying random population oscillations 
rather than strictly periodic temporal evolution; these can be understood as
resonantly amplified demographic fluctuations \cite{at2005}.
Sufficiently large spatially extended predator-prey systems with efficient
predation are similarly characterized by large initial erratic population 
oscillations.
Ultimately, a quasi-stationary state (in the limit of large particle numbers) is 
reached, where both population densities remain non-zero 
\cite{agf1999}--\cite{mmu2007}.
The exceedingly long transients towards this asymptotic predator-prey coexistence 
state are characterized by strong spatio-temporal correlations associated with the
spontaneous formation of spreading activity fronts, depicted in Fig.~\ref{fige1}, 
which induce marked renormalizations of the oscillation parameters as compared to 
the mean-field predictions \cite{miu2006, mmu2007, uct2012}.

To model finite population carrying capacities caused by limited resources,
one may restrict the local particle density or lattice site occupation number
\cite{a1999, ad2000, miu200602, miu2006, mmu2007}.
For the Lotka--Volterra model, this in turn introduces a new absorbing state, 
where the predator population goes extinct, while the prey proliferate through 
the entire system.
By tuning the reaction rates as control parameters, one thus encounters a
continuous active-to-absorbing state non-equilibrium phase transition
\cite{jt1994, nom1994, ad2000, rae2000, ma2001, miu200602, miu2006}. 
In addition, near the predator extinction threshold, population oscillations 
cease, and both predator and prey concentrations directly relax exponentially to
their quasi-stationary values \cite{miu2006}; c.f.~Fig.~\ref{fige2}.

Note that the existence of an absorbing state for the predator population,
which cannot ever recover from extinction under the system's stochastic 
dynamics, explicitly breaks the detailed balance conditions required for systems
to effectively reside in a thermal equilibrium state.
Generically, one expects continuous active-to-absorbing state transitions to be
governed by the universal scaling properties of the directed percolation
universality class (see, e.g., the overviews in Refs.~\cite{noneq1, noneq2, 
uct2014}).
This is in accord with the numerical data obtained in lattice Lotka--Volterra
models; moreover, via representing the corresponding stochastic master equation 
through an equivalent Doi-Peliti pseudo-Hamiltonian and associated 
coherent-state functional integral, one can explicitly map the Lotka--Volterra
model near the predator extinction threshold to Reggeon field theory that 
describes the directed percolation universality class \cite{miu2006, uct2012}.

Induced by severe environmental changes in nature, certain ecosystems may 
collapse and at least some of its species face the danger of extinction. 
In order to monitor viability of populations and maintain ecological diversity,
it is of great importance to identify appropriate statistical indicators that
signify impending population collapse and may thus serve as early warning 
signals. 
In our simulations, we simply perform sudden predation rate switches to mimic 
fast environmental changes. 
If such a rate quench leads near the predator species extinction threshold, we 
observe the characteristic critical slowing-down and aging features expected at
continuous phase transitions \cite{noneq1, noneq2, uct2014}. 
We demonstrate how either of these two characteristic dynamical signatures, but
specifically the emergence of aging scaling, might be utilized as advanced 
warning signals for species extinction \cite{csdexperiment}. 

In the following Sec.~\ref{sec:model-simul-prot}, we describe our stochastic 
lattice Lotka--Volterra model with restricted site occupations and the Monte 
Carlo simulation algorithm.
We then demonstrate in Sec.~\ref{sec:coex_rel} that sudden rate changes within 
the two-species coexistence phase lead to exponentially fast relaxation.
In contrast, Sec.~\ref{sec:crit_rel} explores quenches to the critical predator
extinction threshold, the ensuing critical slowing-down and algebraic predator
density decay, as well as the critical aging scaling that we observe for the 
predator density autocorrelation function.
Finally, Sec.~\ref{sec:concl} provides our conclusions.

\section{Model Description and Simulation Protocol}
\label{sec:model-simul-prot}

We study a spatially extended stochastic Lotka-Volterra model by means of Monte 
Carlo simulations performed on a two-dimensional square lattice with 
$1024 \times 1024$ sites, subject to periodic boundary conditions; we largely
follow the procedures described in Refs.~\cite{miu200602, miu2006, mmu2007}. 
In this work, we impose locally limited carrying capacities for each species
through implementing lattice site occupation restrictions: 
The number of particles per lattice site can only be either $0$ or $1$; i.e., 
each lattice site can either be empty, occupied by a `predator' $A$, or occupied
by a `prey' $B$ particle. 
The individual particles in the system undergo the following stochastic reaction
processes:
\begin{equation}
\label{lvreac}
	A \overset{\widetilde \mu}{\rightarrow} \emptyset \, , \quad
	A + B \overset{\widetilde \lambda}{\rightarrow} A + A \, , \quad 
	B \overset{\widetilde \sigma}{\rightarrow} B + B \, .
\end{equation}

The predators $A$ thus spontaneously die with decay rate $\widetilde \mu > 0$. 
Upon encounter, they may also consume a prey particle $B$ located on a lattice 
site adjacent to theirs, and simultaneously reproduce with `predation' rate 
$\widetilde \lambda > 0$. 
Hence the $B$ particle on the nearest-neighbor site to the predator becomes 
replaced by another $A$ particle.
We remark that in a more realistic description, predation and predator offspring
production should naturally be treated as separate stochastic processes.
While such an explicit separation induces very different dynamical behavior on a 
mean-field rate equation level, it turns out that in dimensions $d < 4$ the 
corresponding stochastic spatially extended system displays qualitatively the very 
same features as the simplified reaction scheme \eref{lvreac} \cite{miu200602}.

Prey in turn may reproduce with birth rate $\widetilde \sigma > 0$, with the 
offspring particles placed on one of their parent's nearest-neighbor sites. 
Note that we do not include nearest-neighbor hopping processes here. 
Instead, diffusive particle spreading is effectively generated through the 
reproduction processes that involve placement of the offspring onto adjacent 
lattice sites; earlier work has ascertained that incorporating hopping 
processes (with rate $\widetilde D$) independent of particle production yields no 
qualitative changes \cite{miu2006}, except for extremely fast diffusion 
$\widetilde D \gg \widetilde \mu, \widetilde \lambda, \widetilde \sigma$, which 
leads to effective homogenization and consequent suppression of spatial 
correlations.
One Monte Carlo Step (MCS) is considered completed when on average all particles
have participated in the above reactions once. 
In our present study, we will hold the rates $\widetilde \mu$ and 
$\widetilde \sigma$ fixed while varying $\widetilde \lambda$ as our control 
parameter.

In general, the detailed Monte Carlo algorithm for the stochastic lattice 
Lotka--Volterra model proceeds as follows \cite{miu2006}:
\begin{itemize}
\item Select a lattice occupant at random and generate a random number $r$ 
      uniformly distributed in the range $[0,1]$ to perform either of the 
      following four possible reactions (with probabilities $D$, $\mu$, 
	  $\lambda$, and $\sigma$ in the range $[0,1]$):
\item If $r < 1/4$, select one of the four sites adjacent to this occupant, and 
      move the occupant there with probability $D$, provided the selected 
      neighboring site is empty (nearest-neighbor hopping).
\item If $1/4 \leq r < 1/2$ and if the occupant is an $A$ particle, then with 
      probability $\mu$ the site will become empty (predator death, 
      $A \rightarrow \emptyset$).
\item If $1/2 \leq r < 3/4$ and if the occupant is an $A$ particle, choose a 
      neighboring site at random; if that selected neighboring site holds a $B$ 
      particle, then with probability $\lambda$ it becomes replaced with an 
      $A$ particle (predation reaction, $A + B \rightarrow A + A$).
\item If $3/4 \leq r < 1$ and if the occupant is a $B$ particle, randomly select
      a neighboring site; if that site is empty, then with probability 
      $\sigma$ a new $B$ particle is placed on this neighboring site (prey 
      offspring production, $B \rightarrow B + B$).
\end{itemize}
To clarify our algorithm, we discuss a simple example: 
If $r$ is generated to be $0.1$, the first case applies, whence we need to 
perform nearest-neighbor hopping with probability $D$. 
Since for simplicity we set $D = 0$ in our simulation, we just skip this step and 
proceed to generate a new random number $r$. 
We remark that different versions of similar simulation algorithms can naturally
be implemented wherein the microscopic reaction processes and their ordering
are varied. 
However, macroscopic long-time simulation results should not qualitatively 
differ for such variations, only the effective reaction rates $\widetilde \mu$, 
$\widetilde \lambda$, $\widetilde \sigma$ in \eref{lvreac}, and the diffusivity 
$\widetilde D$ that result from the microscopic probabilities $\mu$, $\lambda$, 
$\sigma$, and $D$, and the overall time scale will need to be rescaled 
accordingly. 

\begin{figure}[t]
\centering
\includegraphics[width=0.82\columnwidth]{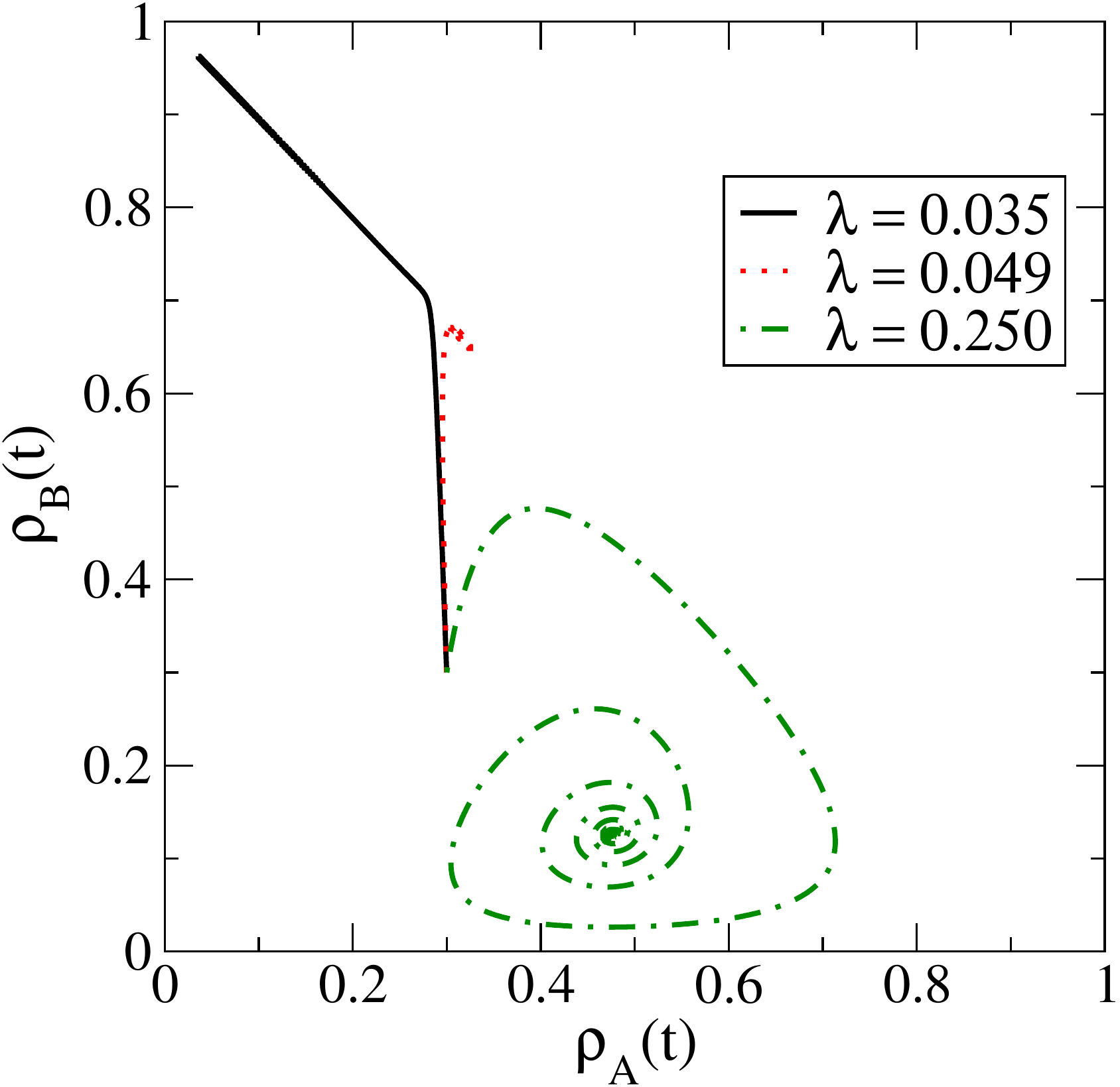} 
\caption{Monte Carlo simulation trajectories for a stochastic Lotka--Volterra 
		model on a $1024 \times 1024$ square lattice with periodic boundary 
		conditions and restricted site occupancy (at most one particle allowed on
		each lattice site) shown in the predator $\rho_A(t)$ versus prey density 
		$\rho_B(t)$ phase plane ($\rho_A + \rho_B \leq 1$) with initial values 
		$\rho_A(0) = 0.3 = \rho_B(0)$, fixed reaction probabilities $\mu = 0.025$,
		$\sigma = 1.0$, and different predation efficiencies:
		(i) $\lambda = 0.035$ (solid black): predator extinction phase;
		(ii) $\lambda = 0.049$ (red dashed): direct exponential relaxation to
		the quasi-stationary state just above the extinction threshold in the 
		predator-prey coexistence phase; 
		(iii) $\lambda = 0.250$ (green dash-dotted): the trajectories spiral into 
		a stable fixed point, signifying damped oscillations deep in the 
		coexistence phase.}
\label{fige2}
\end{figure}

Figures~\ref{fige1} and \ref{fige2} show the results of typical Monte Carlo
simulation runs in two-dimensional stochastic lattice Lotka--Volterra models with
periodic boundary conditions and site occupations restricted to at most a single
predator ($A$) or prey ($B$) particle; thus both their local and mean densities 
are restricted to the range $\rho_A + \rho_B \leq 1$.
Here, $\rho_{A / B}(t)$ is defined as the total predator (prey) number at time 
$t$ divided by the number of lattice sites $(1024)^2$.
The snapshots at various simulation times in Fig.~\ref{fige1} visualize the early
spreading activity fronts in the two-species coexistence phase, i.e., prey (green)
invading empty (white) regions followed by predators (red), which induce the 
initial large-amplitude population oscillations.
At longer run times, the prey here localize into fluctuating clusters, and the
net population densities reach their quasi-stationary values.
In the prey vs. predator density phase plot depicted in Fig.~\ref{fige2}, the
associated trajectory (iii) is a spiral converging to the asymptotic density 
values.
Upon approaching the predator extinction threshold (holding $\mu$ and $\sigma$
fixed, at lower values of $\lambda)$, the population oscillations cease and the
trajectory (ii) relaxes directly to the quasi-stationary coexistence state.
Finally, below the extinction threshold (i), the predator population dies out, and
the prey particles eventually fill the entire lattice.

\section{Results}

\subsection{Relaxation dynamics within the coexistence phase}
\label{sec:coex_rel}

\begin{figure*}[t]
\centering
\includegraphics[width=0.95\columnwidth]{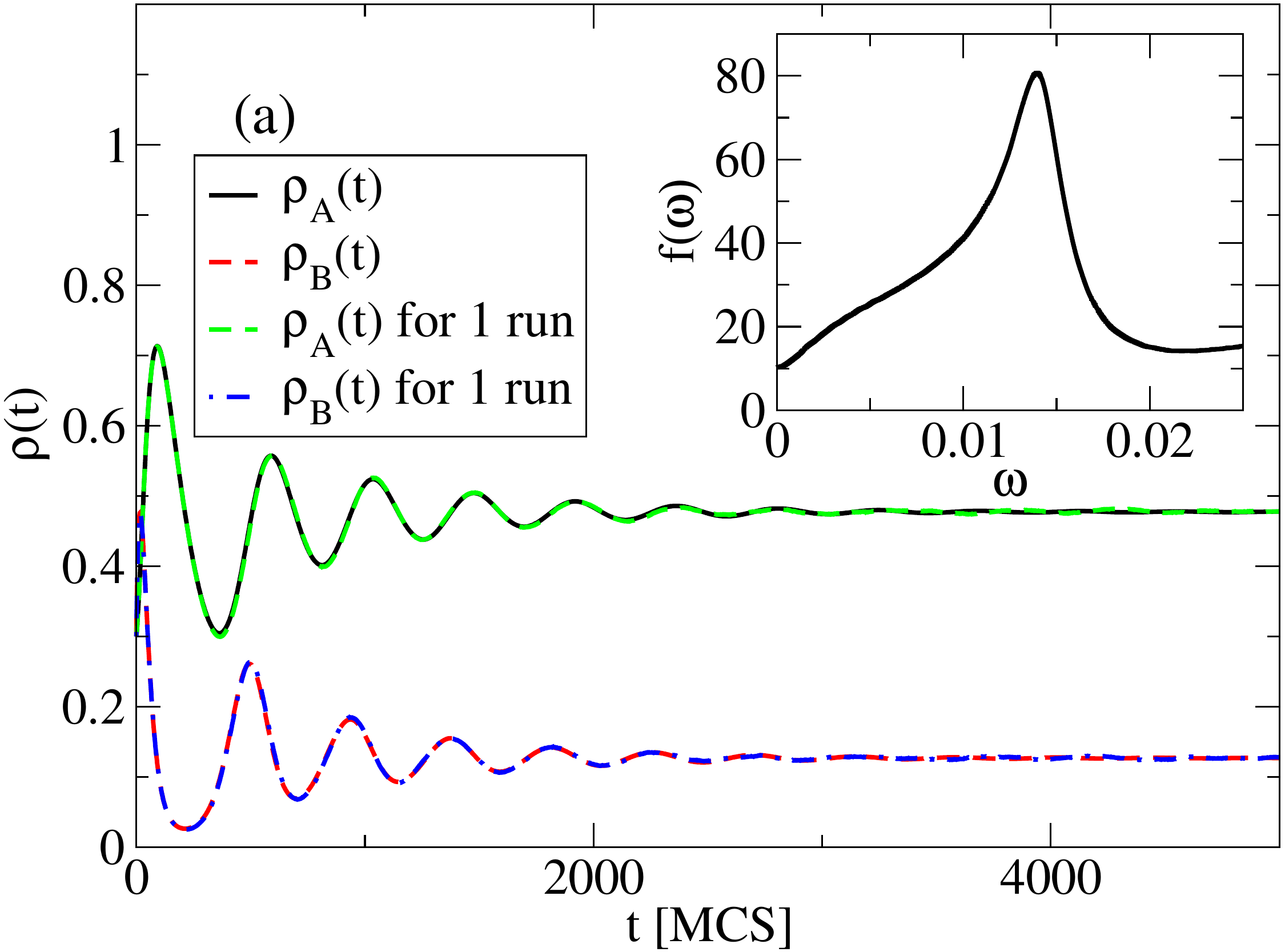} \
\includegraphics[width=1.04\columnwidth]{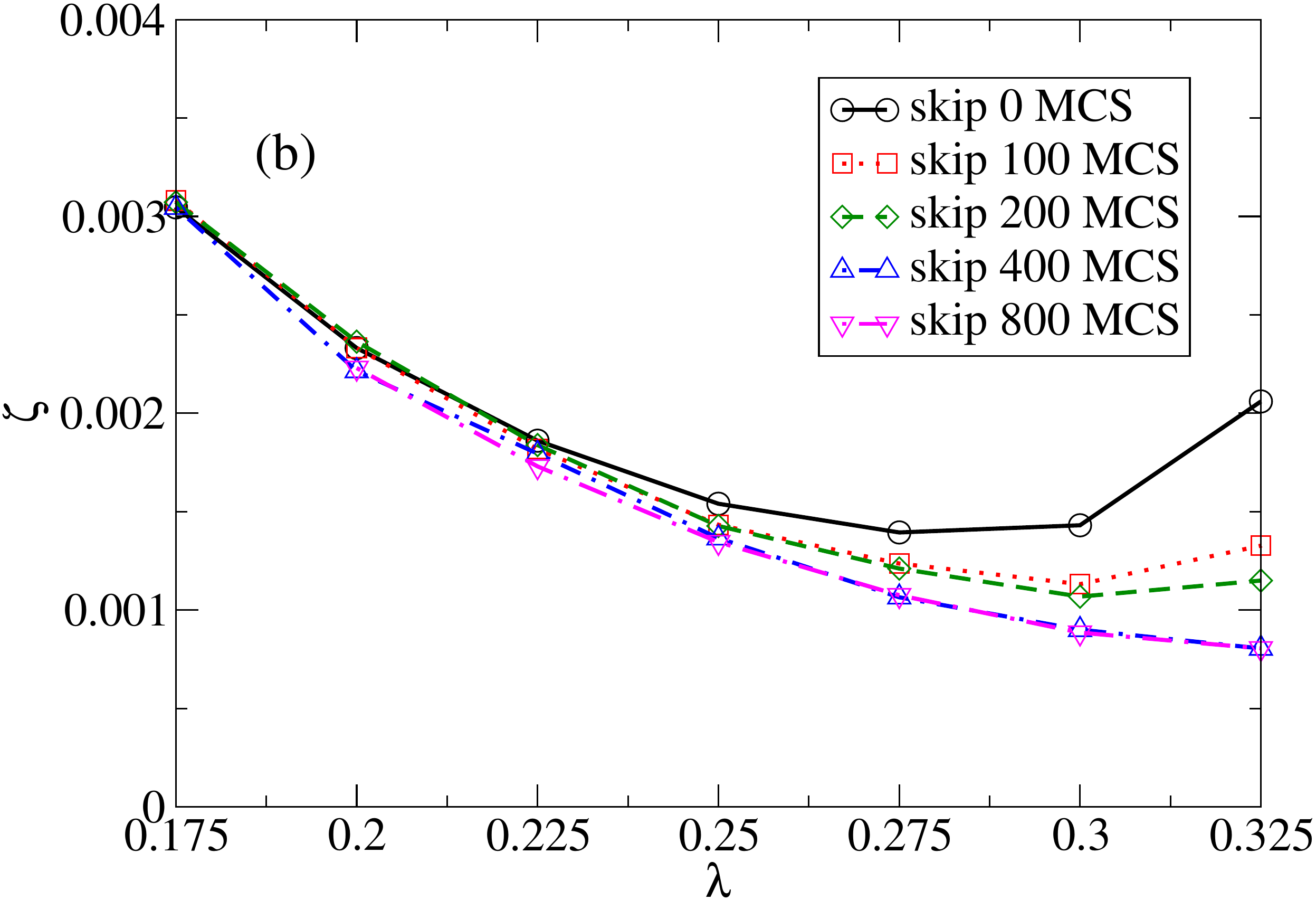}
\caption{Monte Carlo simulation data for a stochastic Lotka--Volterra model on a
	    $1024 \times 1024$ square lattice with periodic boundary conditions:
	      (a) Temporal evolution of the mean predator density $\rho_A(t)$ and the 
                mean prey density $\rho_B(t)$ for fixed reaction probabilities 
		$\mu = 0.025$, $\lambda = 0.25$, and $\sigma = 1.0$, with initial 
		densities $\rho_A(0) = 0.3 = \rho_B(0)$
                (single runs shown in green / blue; the black and red curves are from 
		data averaged over $200$ independent simulation runs). 
	    The inset shows the absolute value of the Fourier transform 
        $f_A(\omega)$ of $\rho_A(t)$ (black curve). 
	    (b) Measured damping rate $\zeta$ in MCS$^{-1}$, as obtained from the 
        peak half-width of $f_A(\omega)$, as function of the predation rate 
        $\lambda$, when respectively $\Delta t = 0$, $100$, $200$, $400$, and 
        $800$ (top to bottom) initial MCS are skipped in the time series 
		$\rho_A(t)$ as the Fourier transform is performed.}
\label{fig1}
\end{figure*}
To set the stage, we first consider the time evolution of our stochastic
Lotka--Volterra system on a two-dimensional lattice starting from a random 
initial configuration with the rate parameters set such that the model resides
within the active coexistence state: the mean densities for both predator and 
prey species will thus remain positive and asymptotically reach constant values. 
Figure~\ref{fig1}(a) shows the time evolution of the mean population densities 
$\rho_A(t)$ and $\rho_B(t)$, averaged over the entire lattice, both for single 
Monte Carlo simulations as well as data that result from averaging over $200$ 
independent runs. 
As is apparent in Fig.~\ref{fige1}, already in moderately large lattices there 
emerge almost independent spatially separated population patches. 
The system is thus effectively self-averaging; as a consequence, the mean data 
from multiple runs essentially coincide with those from single ones.
The reaction rates in the scheme \eref{lvreac} constitute the only control 
parameters in our system. 
Thus, if we fix the probabilities $\sigma$ and $\mu$, the predation probability
$\lambda$ fully determines the final state. 
For generating the data in Fig.~\ref{fig1}, we used constant reaction 
probabilities $\mu = 0.025$, $\lambda = 0.25$, and $\sigma = 1.0$. 
The particles of both species are initially distributed randomly on the lattice
with equal densities $\rho_A(0) = 0.3 = \rho_B(0)$. 

In the early-time regime, the population densities are non-stationary and 
oscillate with an exponentially decreasing amplitude 
$\sim e^{- \zeta t} = e^{- t / t_c}$. 
We may utilize the damping rate $\zeta$ or decay time $t_c = 1 / \zeta$ to 
quantitatively describe the relaxation process towards the quasi-steady state. 
Since $\zeta$ is identical for both species, we just obtain this relaxation rate 
from the predator density decay via measuring the half-peak width of the 
absolute value of the Fourier transform of the time signal,
$f_A(\omega) = \left| \int \rho_A(t) \, e^{i \omega t} \, dt \right|$. 
Alternatively, one might employ a direct fit in the time domain to damped 
oscillations.
However, such a procedure is usually less accurate, as the early-time regime 
tends to be assigned too much weight in determining the fit parameters. 
Using the temporal Fourier transform and deducing the damping rate from the peak 
width in frequency space constitutes a numerically superior method. 
For example, for the predator density $\rho_A(t)$ shown in Fig.~\ref{fig1}(a),
the Fourier transform displayed in the inset yields $t_c \approx 650$ MCS. 
Indeed, by $t = 3000$ MCS $\approx 5\,t_c$, the system has clearly reached its
quasi-stationary state.

In the following, we aim to ascertain that the stochastic lattice 
Lotka--Volterra model loses any memory of its initial configuration once it has
evolved for a duration $t > t_c$. 
We set the initial configuration at $t = 0$ to be a random spatial distribution 
of particles with densities $\rho_A(0) = 0.3 = \rho_B(0)$, and hold 
$\mu = 0.025$ and $\sigma = 1.0$ fixed. 
We then record the relaxation kinetics for various values of $\lambda$, all in
the interval $[0.175, 0.325]$ to ensure that the final states reside deep within
the species coexistence region, and measure the associated damping rates 
$\zeta$.
The full black line in Fig.~\ref{fig1}(b) plots the resulting function
$\zeta(\lambda)$; the relaxation rate is indeed predominantly determined just 
by the reaction rates, but also influenced by the system's initial state.
The initial configuration effects can however be removed in a straightforward
manner as follows:
In the evaluation of the Fourier transform $f_A(\omega)$, we skip a certain 
initial MCS interval $\Delta t$ and just use the remaining data for $\rho_A(t)$ 
rather than the entire time sequence. 
If in fact the initial configuration only affects the system up to time
$t \approx t_c$, the thereby obtained values of the function $\zeta(\lambda)$ 
should become independent of the length of the discarded initial time interval
$\Delta t$ once $\Delta t > t_c(\lambda)$.

\begin{figure*}[t]
\centering
\includegraphics[width=0.96\columnwidth]{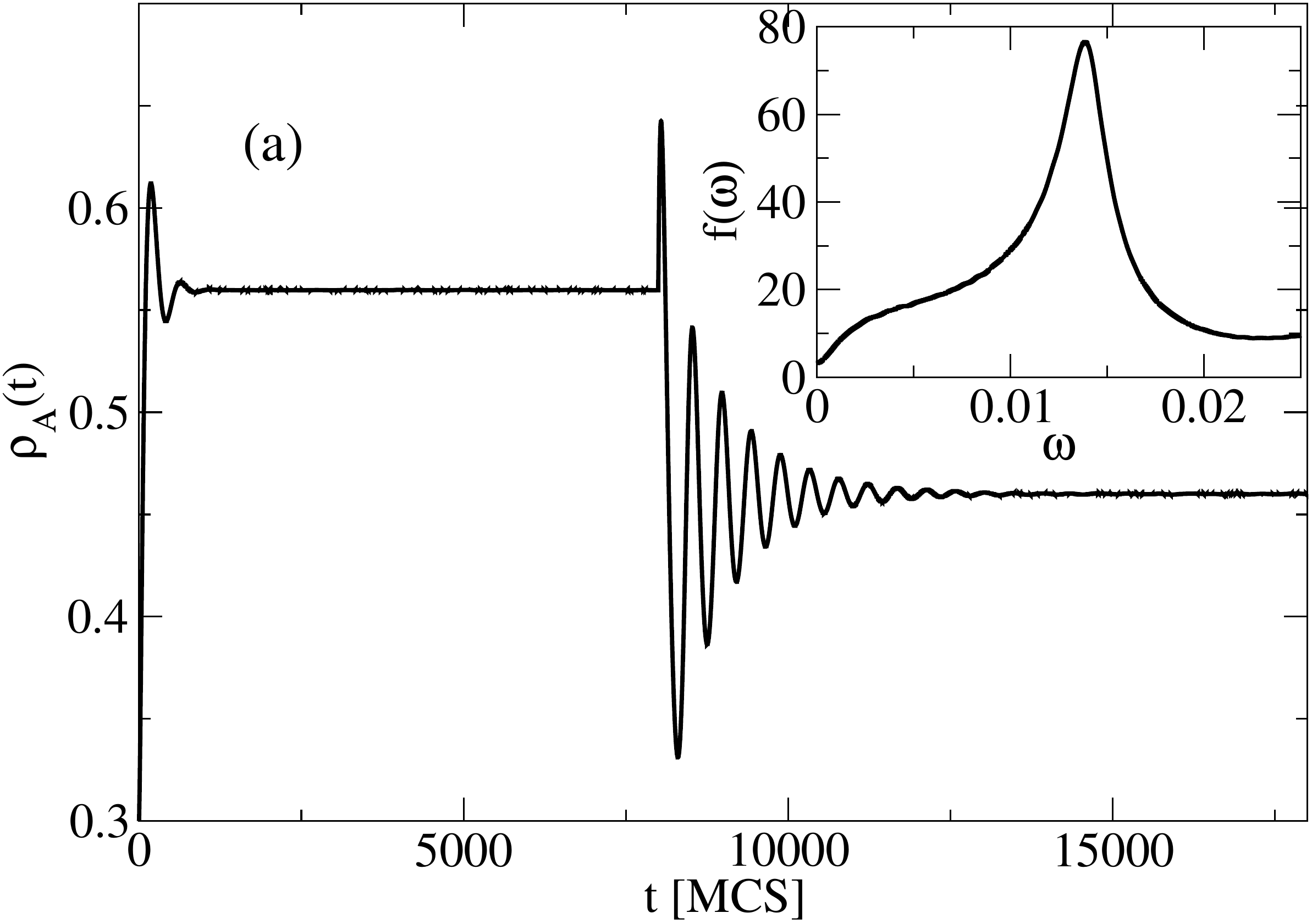} \quad
\includegraphics[width=\columnwidth]{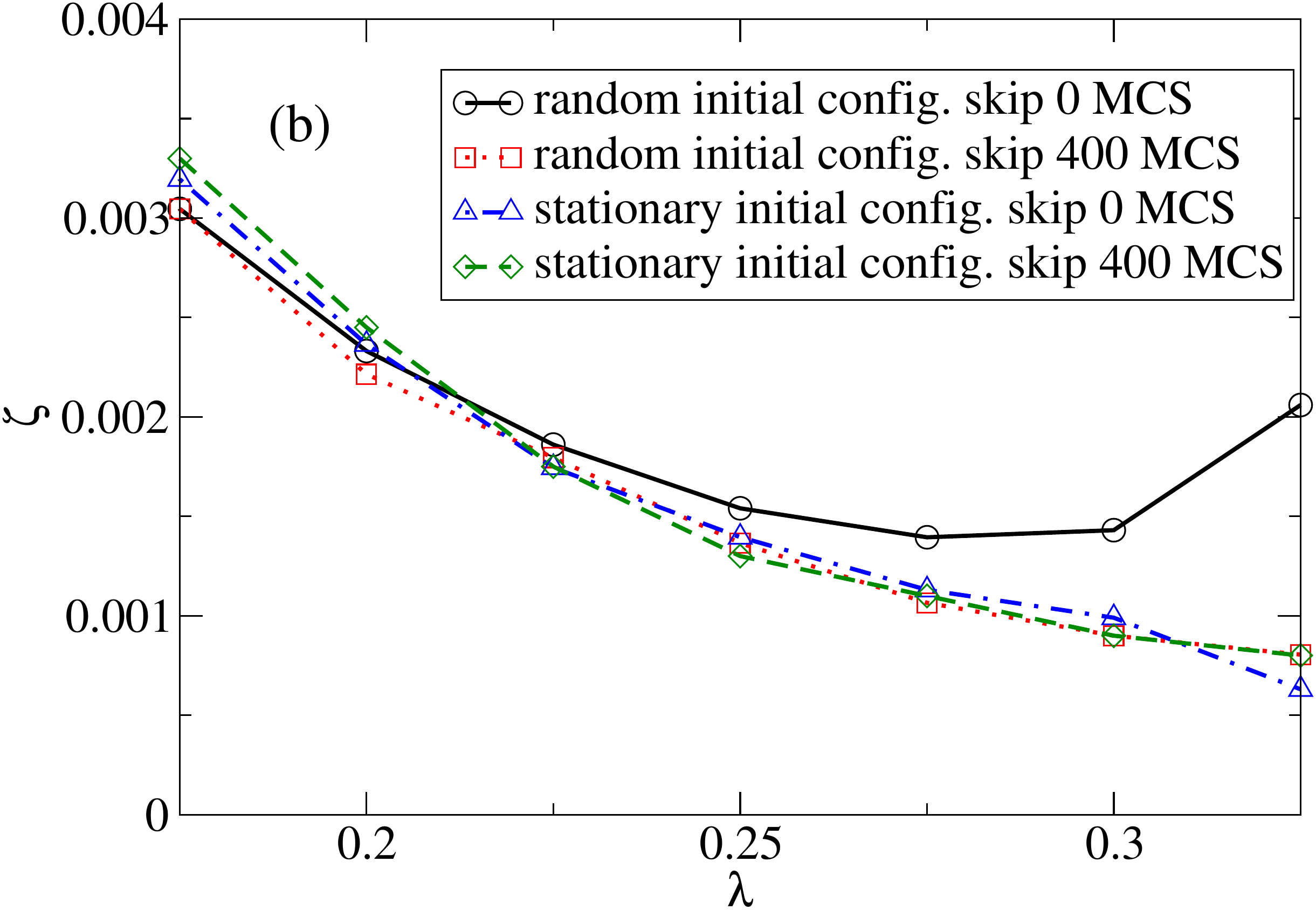}
\caption{Monte Carlo simulation data for the predation rate quench scenario 
	    within the predator-prey coexistence phase on a $1024 \times 1024$
        square lattice with periodic boundary conditions (data averaged over 
        $200$ independent simulation runs): 
	    (a) The main plot shows the temporal evolution of the mean predator 
        density $\rho_A(t)$ for $\mu = 0.025$, $\sigma = 1.0$, 
		$\rho_{A/B}(0) = 0.3$. 
        At $t_1 = 8000$ MCS, the predation probability is suddenly switched from
	    $\lambda_1 = 0.125$ to $\lambda_2 = 0.275$. 
	    The inset displays the absolute value of the Fourier transform 
        $f_A(\omega)$ of the time series after the quench ($t > t_1$).
   	    (b) Measured damping rate $\zeta(\lambda)$ when the system starts from a 
	    random initial condition with skipped initial time interval duration
	    $\Delta t = 0$ (full black line) or $400$ MCS (red dotted line; same data 
		as in Fig.~\ref{fig1}(b)), and following a quench from a quasi-steady 
		state, where $\Delta t = 0$ (blue dash-dotted) and $\Delta t = 400$ MCS 
		(green dashed).}
\label{fig2}	
\end{figure*}
In Fig.~\ref{fig1}(b), we display the functions $\zeta(\lambda)$ obtained for
various skipped initial time interval lengths $\Delta t$, ranging from $0$ to
$800$ MCS.
For large values of the predation probability $\lambda$, and ensuing long 
relaxation times $t_c(\lambda) = 1 / \zeta(\lambda)$, a marked dependence on
$\Delta t$ is apparent.
Yet for the entire $\lambda$ range under investigation, the two curves for
$\Delta t = 400$ MCS and $800$ MCS overlap (and we have checked this holds also 
for other values of $\Delta t > 400$ MCS).
Consequently, any dependencies on the initial configurations in the function 
$\zeta(\lambda)$ have been removed once $\Delta t > 400$ MCS.
This already gives a rough estimate for the typical relaxation rate, 
$\zeta \approx 0.0025$ MCS$^{-1}$, which indeed nicely matches the actual 
values seen in Fig.~\ref{fig1}(b).

Next we check and confirm the above conclusions by considering a set of 
completely different initial configurations: 
Starting again from a random particle distribution with initial particle
densities $\rho_A(0) = 0.3 = \rho_B(0)$, we first let the system relax to its 
quasi-steady state at prescribed reaction probabilities $\mu = 0.025$, 
$\sigma = 1.0$, and $\lambda_1 = 0.125$.
Then we suddenly switch the predation probability to a new value $\lambda_2$.
As a consequence, the system is driven away from the close to stationary, 
spatially highly correlated configurations that are characterized by well-formed 
domains of predators and prey, and relaxes towards a new stable quasi-steady 
state.
Figure~\ref{fig2}(a) illustrates such a `quench' scenario: 
At $t_1 = 8000$ MCS, the predation probability is instantaneously changed from 
the initial value $\lambda_1 = 0.125$ to $\lambda_2 = 0.275$; it is apparent 
that the predation rate switch once again induces large transient population 
oscillations. 
We measure the ensuing relaxation rate $\zeta(\lambda)$ as function of the
post-quench predation probability $\lambda = \lambda_2$ following the procedures
outlined above, and with $\lambda_2$ in the interval $[0.175, 0.325]$.
In Fig.~\ref{fig2}(b) we plot the result, if in the computation of the Fourier
transform $f_A(\omega)$ an initial time interval of duration $\Delta t = 400$ 
MCS is skipped (green dashed curve).
For comparison, we also show the data for $\Delta t = 0$ (blue dash-dotted) and 
replot the corresponding graphs for $\Delta t = 0$ (black line) and 
$\Delta t = 400$ MCS (red dotted) obtained for spatially random initial 
configurations.
Both functions $\zeta(\lambda)$ with $\Delta t = 400$ MCS overlap with the one
for $\Delta t = 0$ initiated in the quasi-steady state:  
Once $\Delta t > t_c(\lambda)$, the system's initial states (here, spatially 
random or correlated) have no noticeable influence on the functional dependence 
of the relaxation rate on $\lambda$.

\subsection{Relaxation kinetics following critical quenches}
\label{sec:crit_rel}

\begin{figure}[t]
\centering
\includegraphics[width=0.98\columnwidth]{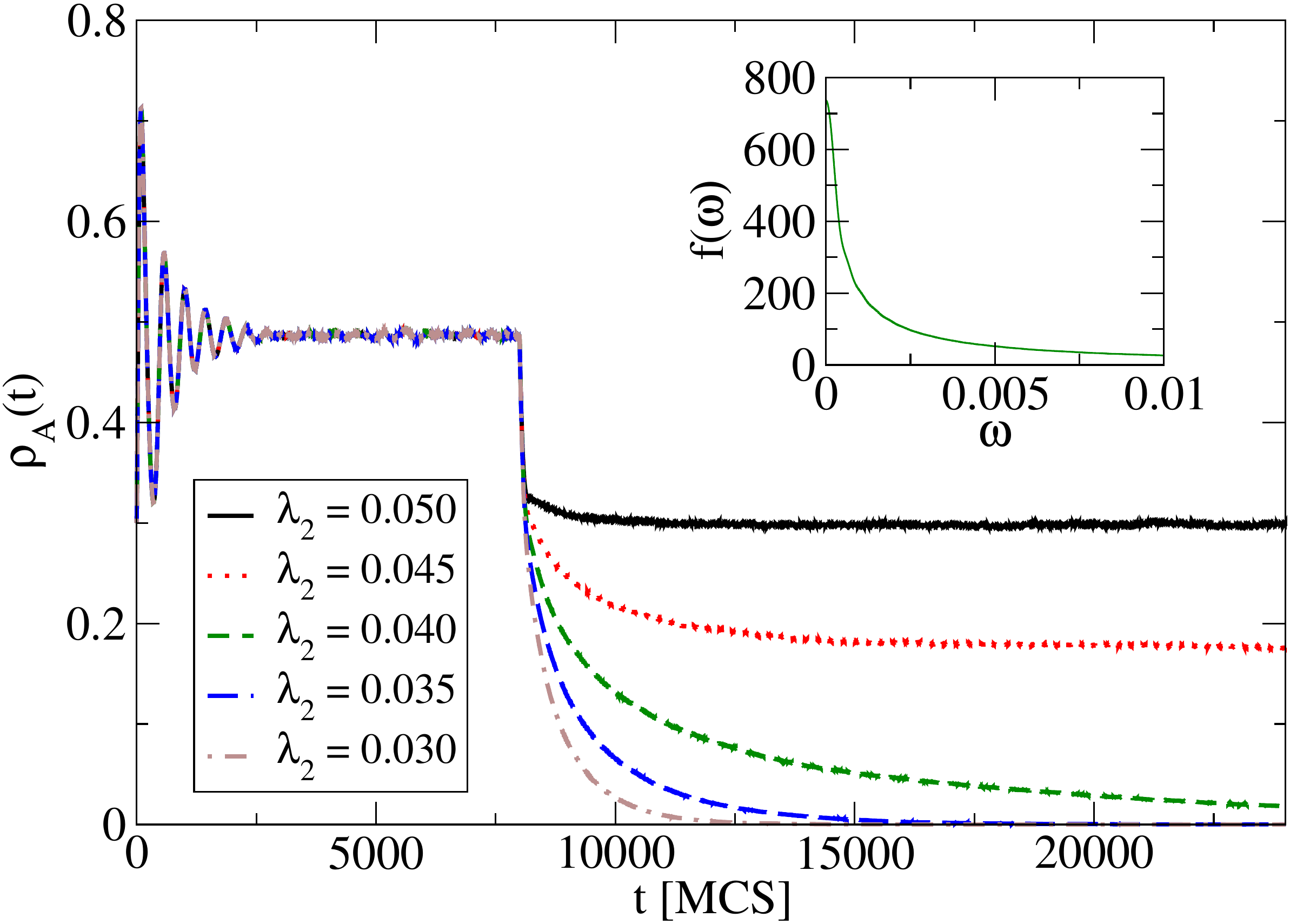}
\caption{Monte Carlo simulation data (single runs) for the mean predator density
	    $\rho_A(t)$ with $\rho_{A/B}(0) = 0.3$ on a $1024 \times 1024$ square 
	    lattice with periodic boundary conditions:	 
	    The initial reaction probabilities are set to $\mu = 0.025$, 
        $\sigma = 1.0$, and $\lambda_1 = 0.250$; at $t_1 = 8000$ MCS the system 
		is quenched into the vicinity of the critical predator extinction point 
		via switching the predation probability to $\lambda_2 = 0.050$, $0.045$, 
        $0.040$, $0.035$, and $0.030$ (top to bottom). 
	    The inset shows the absolute value $f_A(\omega)$ of the Fourier 
        transform of $\rho_A(t)$ for $\lambda_2 = 0.040$ after the quench.}
\label{fig3}
\end{figure}

In the stochastic lattice Lotka--Volterra model with limited local carrying 
capacity, there exists an active-to-absorbing phase transition, namely an 
extinction threshold for the predator population, as illustrated in 
Fig.~\ref{fige2} and Fig.~\ref{fig3}. 
Here, our Monte Carlo simulation runs were initiated with randomly placed 
predator and prey particles with densities $\rho_A(0) = 0.3 = \rho_B(0)$ and 
reaction probabilities $\mu = 0.025$, $\sigma = 1.0$, and $\lambda_1 = 0.250$. 
After $t_1 = 8000$ MCS, when the system has clearly reached its quasi-steady 
state, we suddenly switch the predation probability from $\lambda_1$ to much 
lower values $\lambda_2$ in the range $[0.03, 0.05]$, which reside in the 
vicinity of the extinction critical point. 
The curves in Fig.~\ref{fig3} show the ensuing time evolution for the mean 
predator density $\rho_A(t)$. 
For small $\lambda_2$, all predator particles disappear, and eventually only the
prey species survives, in contrast to the active two-species coexistence phase
at larger $\lambda_2$ values, where both species persist with finite, stable 
mean population densities. 
In this subsection, we investigate in detail the non-equilibrium relaxation
properties and ensuing dynamic scaling behavior following the quench from a 
quasi-steady coexistence state to the critical point. 

\begin{table}[b]
\begin{center}
\begin{tabular}{|c|c|c|c|}
      \hline $d = 2$ & SLLVM & DP sim. & DP exp. \\ \hline 
      $\alpha$ & 0.540(7) & 0.4505(10) \cite{612} & 0.48(5) \cite{experiment} \\ 
      $z\, \nu$ & 1.208(167) & 1.2950(60) \cite{249} & 1.29(11) \cite{experiment}
      \\ \hline
      $\Lambda_c/ z$ & 2.37(19) & 2.8(3) \cite{609} & 2.5(1) \cite{experiment}\\ 
      $b$ & 0.879(5) & 0.901(2) \cite{609} & 0.9(1) \cite{experiment} \\ \hline
\end{tabular}
\end{center}
\caption{Monte Carlo simulation results for the critical exponents near the 
	    predator extinction threshold in the stochastic lattice Lotka--Volterra 
	    model (SLLVM) on a square lattice. 
	    For comparison, the table also lists the accepted literature scaling 
	    exponents for the directed percolation (DP) universality class 
        \cite{612, 249} as well as experimental values measured in turbulent 
        liquid crystals (MBBA)	\cite{experiment}.
	    The critical aging scaling exponents as obtained in our simulations 
        along with earlier numerical results for the contact process \cite{609} 
        and experimental data \cite{experiment} are included as well. 
	    The numbers in brackets indicate the estimated uncertainty in the last 
	    digits.}
\label{table}
\end{table}
Near the active-to-absorbing phase transition, in our two-dimensional 
stochastic Lotka--Volterra model located at $\lambda_c = 0.0416 \pm 0.0001$ for 
fixed $\mu = 0.025$ and $\sigma = 1.0$, one should anticipate the standard 
critical dynamics phenomenology for continuous phase transitions 
\cite{noneq1, uct2014}:
Fluctuations become prominent, and increasingly large spatial regions behave
cooperatively, as indicated by a diverging correlation length 
$\xi(\tau) \sim |\tau|^{-\nu}$, where $\tau = (\lambda / \lambda_c) - 1$.
Consequently, the characteristic relaxation time should scale as 
\begin{equation}
\label{crslow}
	t_c(\tau) \sim \xi(\tau)^z \sim |\tau|^{-z \, \nu} \ , 
\end{equation}
implying a drastic critical slowing-down of the associated dynamical processes.
Thus, exponential relaxation with time becomes replaced by much slower algebraic
decay of the predator population density precisely at its extinction threshold
$\lambda_c$,
\begin{equation}
\label{cridec}
	\tau = 0 : \ \rho_A(t) \sim t^{-\alpha} \ .
\end{equation}	
The values of the three independent critical scaling exponents $\nu$, $z$, and 
$\alpha$ characterize certain dynamical universality classes.
Generically, one expects active-to-absorbing phase transitions to be governed by
the critical exponents of directed percolation (DP) \cite{noneq1, uct2014}; the
middle column in Table~\ref{table} lists the established numerical literature 
values (from Refs.~\cite{612, 249}) for $\alpha$ and the product $z \, \nu$ in 
two dimensions.
Indeed, standard field-theoretic procedures allow a mapping of the near-critical
stochastic spatial Lotka--Volterra system to Reggeon field theory 
\cite{miu2006, uct2012}, which represents the effective field theory for 
critical directed percolation \cite{hjuct2005, uct2014}.

\begin{figure}
\centering
\includegraphics[width=0.98\columnwidth]{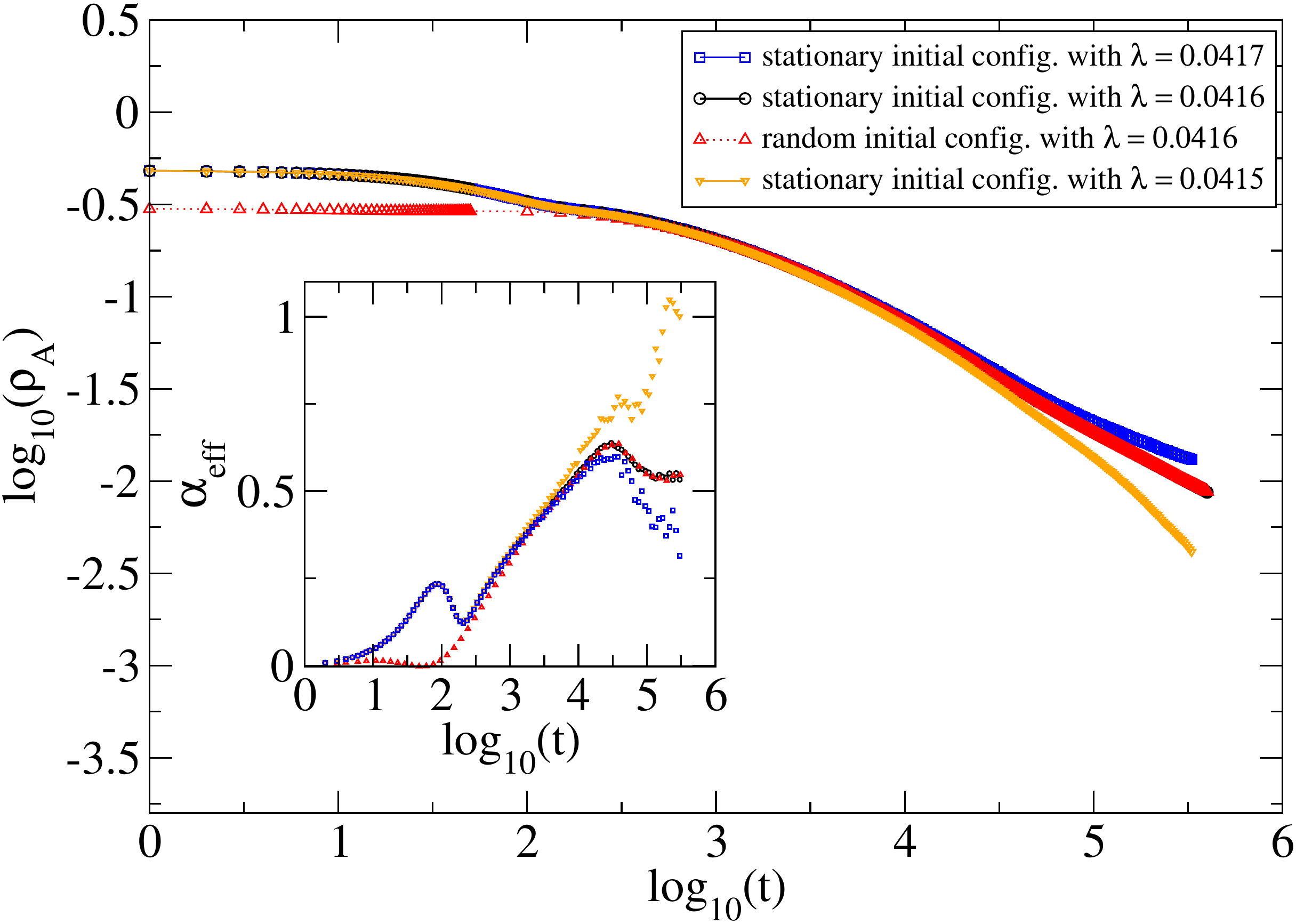}
\caption{Double-logarithmic plot of the mean predator density decay $\rho_A(t)$ 
        at the critical extinction threshold $\lambda_c = 0.0416$ for fixed
		$\mu = 0.025$ and $\sigma = 1.0$, for both quasi-stationary (top, black)
        and random (lower curve, red dotted) initial configurations (data 
        averaged over $2000$ independent simulation runs).
		For comparison, the predator density decay data are shown also for 
		$\lambda = 0.0417$ (blue, active coexistence phase) and 
		$\lambda = 0.0415$ (orange, absorbing predator extinction phase).
		The graphs in the inset show the local negative slopes of the curves in 
        the main panel (using intervals of size $0.05$), i.e., the 
        time-dependent effective critical decay exponent $\alpha_{\rm eff}(t)$.}
\label{fig4}
\end{figure}
In order to analyze the dynamic critical scaling behavior at the predator
population extinction threshold, we follow the protocol outlined above and shown
in Fig.~\ref{fig3}, and first let the system relax to its quasi-steady state 
with a large predation probability $\lambda_1 = 0.250$.
After $8000$ MCS, we quench the system to its critical point by switching this
probability instantaneously to $\lambda_c = 0.0416$. 
In order to acquire decent statistics, we perform $2000$ independent Monte Carlo
simulation runs and then average our results. 
Figure~\ref{fig4} shows a double-logarithmic plot of the decay of the mean 
predator density $\rho_A(t)$ following the quench (top black curve); the inset 
displays the measured local slope of this graph (taken with intervals of size 
$0.05$), which can be interpreted as a time-dependent effective critical decay 
exponent $\alpha_{\rm eff}(t) = - d \log \rho_A(t) / d \log t$. 
A clean power law decay is thus reached when the slope stays constant, which in 
our data happens only quite late, at $t > 10^5$ MCS. 
An asymptotically constant slope is observed as long as the value of $\lambda_2$
is sufficiently close to the critical point $\lambda_c$. 
Our simulation data (in the rather short time interval with constant 
$\alpha_{\rm eff}$) yield the critical decay exponent value 
$\alpha = 0.54 \pm 0.007$; we note that the accepted directed percolation 
universality class value is $\alpha \approx 0.4505$, which has been obtained by 
performing activity spreading simulations \cite{612}, see also 
Ref.~\cite{miu2006}. 
For comparison, we also display the data for quenches to $\lambda = 0.0417$ (blue
curve) and $\lambda = 0.0415$ (orange): 
In the former case, the predator density approaches a finite value as the system 
is still, albeit barely, in the active two-species coexistence phase, while for
the lower predation rate the absorbing state is reached, with the predator
population going extinct.

We wish to ascertain the independence of the asymptotic critical exponent
$\alpha = \alpha_{\rm eff}(t \to \infty)$ from the starting configurations in
our simulations. 
To this end, we directly initialize our stochastic lattice Lotka--Volterra
system with reaction probabilities $\mu = 0.025$, $\sigma = 1.0$, and 
$\lambda = \lambda_c = 0.0416$.
The resulting Monte Carlo data, averaged over again $2000$ independent runs, 
are also shown in Fig.~\ref{fig4} (lower, dotted red curve).
It is apparent that for the selected parameter values, the initial conditions
become irrelevant for $t > 1000$ MCS, whereafter our results for random and 
correlated quasi-steady state initial conditions perfectly overlap.

\begin{figure}
\centering
\includegraphics[width=0.95\columnwidth]{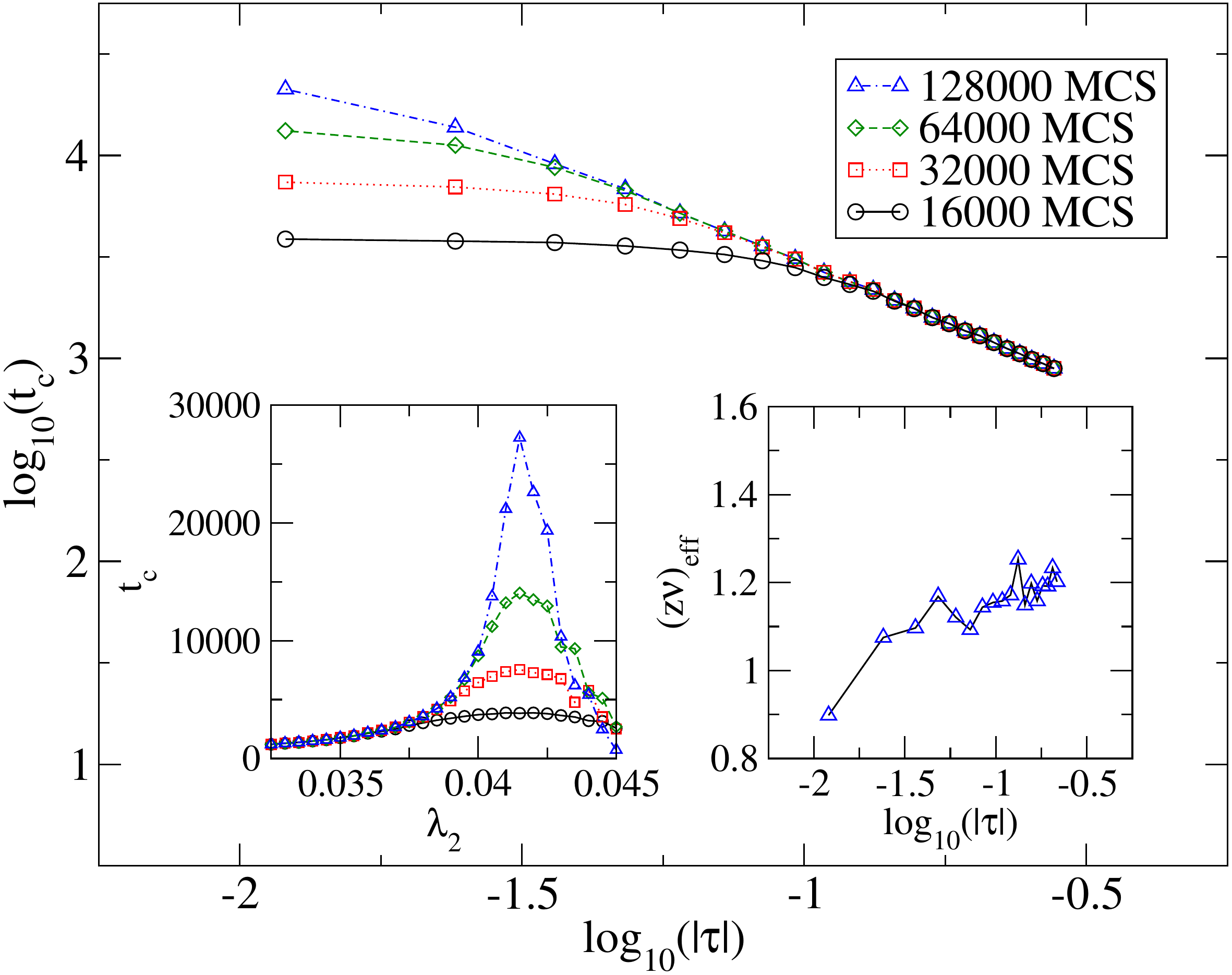}
\caption{Characteristic relaxation time $t_c$ for the near-critical stochastic 
        Lotka--Volterra model on a $1024 \times 1024$ square lattice.
        The left inset shows the relaxation time $t_c(\lambda_2)$ when the system 
		is quenched from a quasi-stationary state at $\lambda_1 = 0.25$ to 
		smaller values $\lambda_2$ in the vicinity of the predator extinction 
		threshold $\lambda_c = 0.0416$. 
        The different curves indicate the resulting relaxation time estimates 
        when respectively $128000$, $64000$, $32000$, and $16000$ MCS (top to 
        bottom) were performed after the quench (all data averaged over $500$ 
        independent simulation runs).
        The main panel displays the same data in double-logarithmic form.
        For $|\tau| = |(\lambda_2 / \lambda_c) - 1| > 0.1$, the different graphs 
        overlap, collapsing onto a straight line with slope 
        $-z \, \nu = -1.208 \pm 0.167$.
		The right inset shows the associated local slope or effective exponent
		$(z \, \nu)_{\rm eff}(t)$ that appears to tend towards 
		$z \, \nu \approx 1.3$.}
\label{fig5}
\end{figure}

As visible in Fig.~\ref{fig3}, the stochastic spatial Lotka--Volterra system 
immediately senses the predation rate change and the predator density decreases 
rapidly after the quench; for a very brief time period (about $200$ MCS), this 
decay is in fact independent of the new predation probability $\lambda_2$.
Subsequently, the relaxation curves are quite sensitive to the selected value 
of $\lambda_2$.
We measure the ensuing relaxation times $t_c(\lambda_2)$ through evaluating the
temporal Fourier transform, while skipping an initial time interval of duration
$\Delta t = 200$ MCS right after the quench in accord with the previous
observation, averaging over $500$ independent Monte Carlo simulation runs. 
Figure~\ref{fig5} depicts our numerically determined relaxation time data for
$t_c$ as a function of $\lambda_2$ in the vicinity of the critical predator
extinction threshold $\lambda_c$.
Note in the left inset that since we only consider a finite time interval (at 
most $128000$ MCS) after the quench for the computation of $f_A(\omega)$, our 
thus measured relaxation times do not diverge, but display a peak as function of 
$\lambda_2$ that becomes both more pronounced and sharper as the duration of the 
Monte Carlo simulation is increased.
Indeed, we can estimate the critical value $\lambda_c$ from the peak position of
the data curve for the longest simulations runs.  

Since the left part of the graph turns out considerably smoother than the right
part, we perform a power law fit only to the corresponding data subset 
($\tau < 0$).
As shown in the main panel of Fig.~\ref{fig5}, this yields the critical 
exponent product $z \, \nu = 1.208 \pm 0.167$; within our statistical error 
bars this is in reasonable agreement with the two-dimensional directed 
percolation literature values $\nu \approx 0.7333$ and $z = 1.7660$ 
\cite{noneq1}, i.e., $z \, \nu \approx 1.295$ \cite{249}.
The right inset displays the local time-dependent effective exponent
$(z \, \nu)_{\rm eff}(t) = - d \log t_c(\tau) / d \log |\tau|$ for the longest
simulation runs; the data appear not settled at a constant value yet, but indeed 
rather tend towards the slightly larger asymptotic DP value. 
Even with our longest Monte Carlo simulation runs, we have at best just barely
reached the asymptotic scaling regime, and thus cannot very precisely determine
the associated critical exponents for quasi-stationary observables.

A remarkable experiment with yeast cells has recently demonstrated a drastic
increase in the relaxation time of the dynamics for a biological system near 
its population extinction threshold \cite{csdexperiment}. 
The authors therefore propose to utilize critical slowing-down as an indicator
to provide advanced warning of an impending catastrophic population collapse. 
However, as we have demonstrated in our simulations, c.f. Figs.~\ref{fig4} and
\ref{fig5}, the unique and universal power law features in the population
density decay and divergence of the relaxation time are asymptotic phenomena 
and emerge only rather late, in our system after $t \approx 10^5$ MCS.
Such a long required time period to unambiguously confirm an ecological 
system's proximity to an irreversible tipping point may preclude timely
interventions to save the endangered ecosystem. 
As we shall see in the following, the appearance of physical aging features in 
appropriate two-time observables may serve as more advantageous warning signals 
for population collapse, as they provide both earlier and often more accurate 
indications for critical behavior (see, e.g., Ref.~\cite{gduct2012}).  

As a consequence of the drastic slowing-down of relaxation processes, 
near-critical systems hardly ever reach stationarity.
During an extended transient period, time translation invariance is broken, and 
the initial configuration strongly influences the system: 
both these features characterize the phenomenon of physical aging \cite{noneq2}. 
In the vicinity of critical points as well as in a variety of other instances
where algebraic growth and decay laws are prominent, the non-equilibrium aging 
kinetics is moreover governed by dynamical scaling laws.
In our context, aging scaling is conveniently probed in the two-time particle 
density autocorrelation function 
\begin{equation}
\label{dencor}
	C(t,s) = \langle n_i(t) \, n_i(s) \rangle - 
	\langle n_i(t) \rangle \, \langle n_i(s) \rangle \ , 
\end{equation}
where $n_i(t) = 0$ or $1$ 
indicates the occupation number (here, for the predators $A$) on lattice site 
$i$ at time $t$. 
The cumulant \eref{dencor} thus measures local temporal correlations as function
of the two time instants $s$ and $t > s$; we shall refer to $t$ and $s$ as the 
observation and waiting time, respectively.

In a stationary dynamical regime, reached for both $t,s > t_c$, time translation 
invariance should hold, implying that $C(t,s)$ becomes a function of the evolved
time difference $t - s$ only.
In the transient aging window, characterized by double time scale separation
$t_c \gg t, s, t - s \gg t_{\rm mic}$, where $t_{\rm mic}$ represents typical 
microscopic time scales, one often encounters a simple aging scenario described
by the dynamical scaling form 
\begin{equation}
\label{agscal}
	C(t,s) = s^{-b} \, f_c(t/s) \, , \quad f_c(y) \sim y^{-\Lambda_c / z}
\end{equation}
with the aging scaling exponent $b$, and a scaling function $f_c$ that 
asymptotic obeys a power law decay in the long-time limit, governed by the ratio
of the autocorrelation exponent $\Lambda_c$ and dynamic critical exponent $z$ 
\cite{noneq2}. 
Near continuous phase transitions both in and far from thermal equilibrium, the
simple aging dynamical scaling form \eref{agscal} can be derived by means of
renormalization group methods \cite{hjuct2005, uct2014}.
For directed percolation, the aging exponents are related to the 
quasi-stationary critical exponents through the scaling relations (in $d$ space
dimensions) \cite{noneq2}
\begin{equation}
\label{scarel}
	b = 2 \, \alpha \ , \quad \Lambda_c / z = 1 + \alpha + d / z \ .
\end{equation}
In perhaps the simplest lattice realization of the directed percolation
universality class, the contact process, numerical simulations have confirmed
the simple aging scaling form \eref{agscal}, with scaling exponents 
$b \approx 0.9$ and $\Lambda_c / z \approx 2.8$ in two dimensions, see 
Table~\ref{table} \cite{609}.
If universality holds near active-to-absorbing phase transitions, we should
observe the same scaling properties at the predator extinction threshold in our
stochastic lattice Lotka--Volterra model.

\begin{figure*}
\centering
\includegraphics[width=\columnwidth]{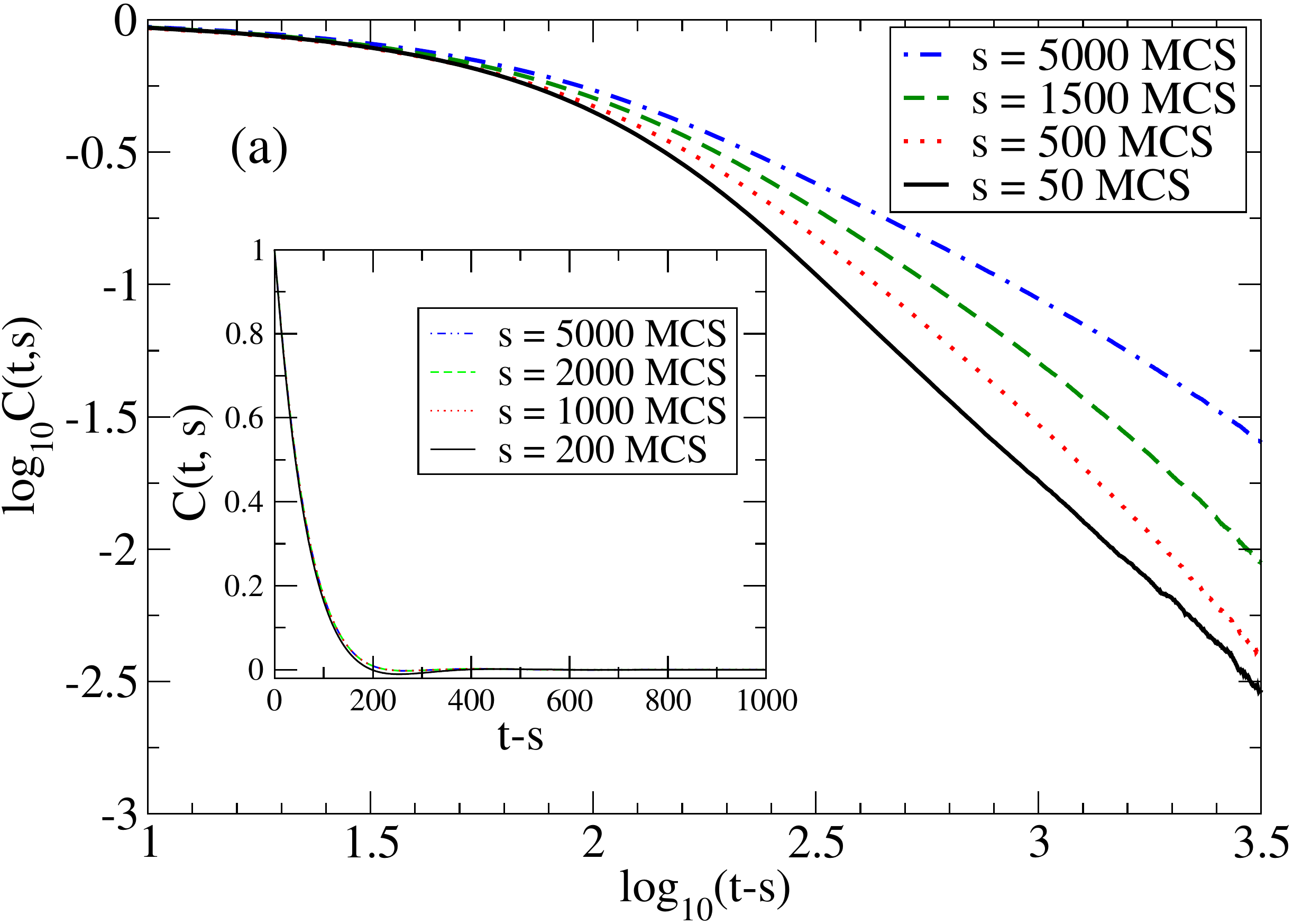} \;\
\includegraphics[width=0.98\columnwidth]{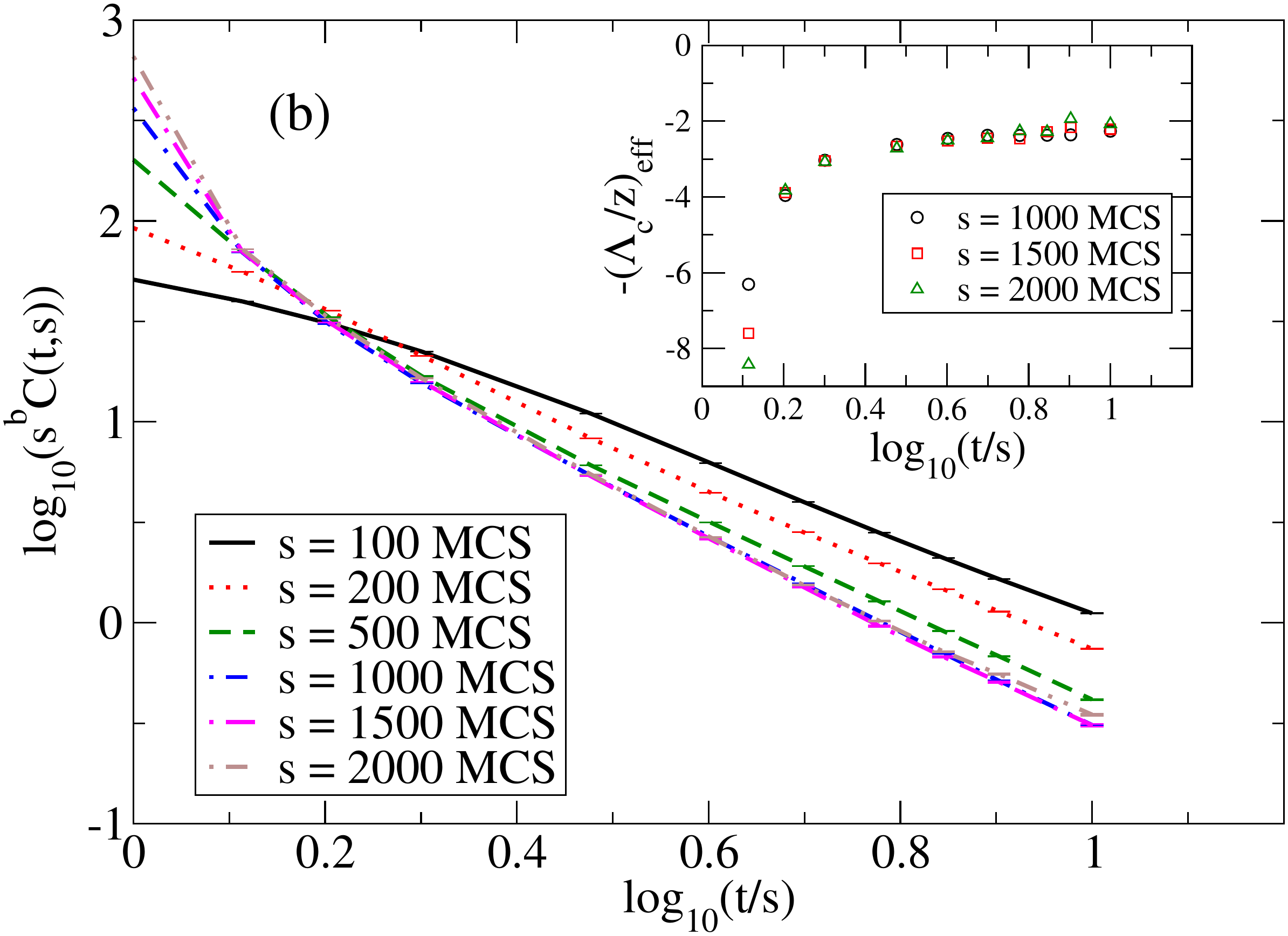}
\caption{(a) Double-logarithmic plot of the predator density autocorrelation 
		function $C(t,s)$ as a function of the time difference $t - s$ for 
		various waiting times $s = 50, 500, 1500, 5000$ MCS (left to right) at 
		the predator extinction critical point $\lambda_c = 0.0416$; the data are 
		averaged over $100$ independent simulation runs for each value of $s$.
		The inset shows that in contrast $C(t,s)$ decays exponentially fast for 
		$\lambda_2 = 0.125$, i.e., a quench within the coexistence phase, and
		demonstrates that time translation invariance holds in this situation
		(data averaged over $400$ simulation runs for the waiting times 
		$s = 5000, 2000, 1000$, and $200$ (top to bottom).
		(b) Simple aging dynamical scaling analysis: $s^b \, C(t,s)$ is graphed 
		versus the time ratio $t / s$, for $1000$ independent simulation runs for 
		each waiting time $s$.
        The straight-slope section of the curves with large waiting times 
		$s \geq 1000$ MCS yields $\Lambda_c / z =  2.37 \pm 0.19$, and the aging 
		scaling exponent is determined to be $b = 0.879 \pm 0.005$.
		The inset displays the local effective exponent 
		$-(\Lambda_c / z)_{\rm eff}(t)$.}
\label{fig6}	
\end{figure*}
In our Monte Carlo simulations, the predator density autocorrelation function
$C(t,s)$ is obtained in a straightforward manner by monitoring the occupations
of the lattice sites with $A$ particles following the previously discussed 
predation probability quench scenario to its critical value $\lambda_c$ after
the system had first reached a quasi-stationary state. 
If time translation invariance applies, the ensuing curves of $C(t,s)$ plotted 
against $t - s$ should overlap for different values of the waiting time $s$. 
Indeed, this is clearly seen in the inset of Fig.~\ref{fig6}(a) for the expected 
exponentially fast relaxation of the predator density autocorrelation function 
following a sudden quench from $\lambda_1 = 0.250$ to $\lambda_2 = 0.125$, whence
the system remains within the two-species coexistence phase.
In stark contrast, as becomes apparent in Fig.~\ref{fig6}(a), time translation 
invariance is manifestly broken at the critical point. 

In Fig.~\ref{fig6}(b), we plot our data for the critical density 
autocorrelations, now averaged over $1000$ simulation runs, in the form of 
$s^b \, C(t,s)$ versus the time ratio $t / s$, for a set of waiting times 
$100$ MCS $\leq s \leq 2000$ MCS, in order to test for the simple aging 
dynamical scaling scenario \eref{agscal}.
The aging exponent $b$ is determined by attempting to collapse the data for the
three large waiting times $s \geq 1000$~MCS onto a single master curve.
With the choice $b = 0.879 \pm 0.005$, the predator density autocorrelation 
function displays simple aging scaling for $s = 1000$, $1500$, and $2000$ MCS.
However, for small $s \leq 500$~MCS, the curves cannot be properly rescaled 
and collapsed.  
Depending on how many data points in the $t / s$ plot are used, one obtains
slightly different values for $b$; their standard deviation gives our estimated 
errors.
Within these error bars, the directed percolation scaling relation
$b = 2 \, \alpha$ is just marginally fulfilled.
We also remark that since our estimate for the location of the critical point
$\lambda_c$ is inevitably measured only with limited accuracy, and we cannot
meaningfully extend our finite-system simulation runs for arbitrarily long time 
periods, we also do not observe aging scaling anymore for $s \gg 2000$ MCS; to 
extend the aging analysis to larger waiting times would require both a more 
accurate measurement of $\lambda_c$ and larger simulation domains.

Finally, the exponent ratio $\Lambda_c / z$ may be estimated from the slope of
the master curve in Fig.~\ref{fig6}(b) resulting from the data collapse at large
waiting times; see also the inset that shows the associated effective exponent.
We find $\Lambda_c / z =  2.37 \pm 0.19$, which within the estimated error bars 
is in fair agreement with the value $2.8$ measured for the contact process in 
$d = 2$ dimensions \cite{609}, and in accord with the scaling relation 
\eref{scarel}. 
To ascertain that the critical aging exponents do not depend on the initial 
configurations, we have repeated the above procedures for Monte Carlo 
simulations initiated with randomly distributed particles at the critical
predation probability $\lambda_c$ rather than starting with the spatially 
correlated initial configurations prepared in quasi-steady states.
We have confirmed that we thereby obtain identical values for $b$ and 
$\Lambda_c / z$, as listed in the first column of Table~\ref{table}.
Within our error bars these are in accord with the two-dimensional values 
obtained for the contact process \cite{609} and in experiments on turbulent
liquid crystals \cite{experiment}.

Figure~\ref{fig6}(b) demonstrates that convincing critical aging scaling 
collapse is achieved for $t / s \geq 5$, or $t \approx 10^4$ MCS even for the 
longest waiting time $s = 2000$ MCS under consideration here.
Note that aging scaling thus appears by about a factor of $10$ earlier in the
system's temporal evolution than the critical power laws describing the 
quasi-stationary predator density decay and the divergence of the characteristic 
relaxation time. 
Just as critical slowing-down, the emergence of physical aging and certainly the
associated dynamical scaling is an unambiguous indicator of the ecosystem's
proximity to population extinction. 
Critical aging scaling hence provides a complementary warning signal for
impending collapse, yet becomes manifestly visible markedly earlier in the 
system's time evolution. 

We end our discussion of non-equilibrium relaxation processes in spatially
extended stochastic Lotka--Volterra models by briefly addressing the sole
remaining quench scenario, which takes the system from the active two-species 
coexistence state to the absorbing phase wherein the predator population goes 
extinct.
Outside the critical parameter region, the characteristic relaxation time $t_c$
is finite; i.e., the mean predator density will decay to zero exponentially 
fast, as local predator clusters become increasingly dilute, while the prey 
gradually fill the entire system.
In the absorbing state, population density fluctuations eventually cease, 
whence no interesting dynamical features remain.

\section{Conclusion}
\label{sec:concl}

To conclude, we have investigated non-equilibrium relaxation features in a
stochastic Lotka--Volterra Model on a two-dimensional lattice via detailed Monte 
Carlo simulations.
If the prey carrying capacity is limited, i.e., in the presence of site 
restrictions (and for sufficiently large system size), there appears a predator 
extinction threshold that separates an inactive phase wherein the prey 
proliferate and the predators die out, from an active phase where both species
coexist and compete. 
In a first set of numerical experiments, we observe the system's relaxation 
either from a random initial configuration or between two quasi-stationary 
states within the active coexistence phase via suddenly changing the predation 
rate. 
As expected, we find that the initial state generically only influences the 
subsequent oscillatory dynamics for the duration of about one characteristic
relaxation time, implying that the system exponentially quickly loses any memory
of the initial configuration. 

Our main focus has thus been the analysis of critical quenches and the ensuing
dynamical scaling behavior.
Following a quench of the predation rate to its critical value for the predator 
species extinction threshold, we have measured the dynamic scaling exponents for
the diverging relaxation time and the algebraic decay of the predator density.
Within our systematic and statistical errors, we obtained the expected values 
for the directed percolation universality class that generically characterizes
active-to-absorbing phase transitions. 
In addition, we have studied the critical aging properties of this system:
Reflecting critical slowing-down, the characteristic relaxation time diverges at
the extinction threshold; as a consequence, time translation invariance is 
broken, and physical aging governed by universal scaling features emerges.
Our measured aging scaling exponents are close to those found previously for
the contact process, which is perhaps the simplest lattice realization of the 
directed percolation universality class.
We remark that at least to our knowledge, this present study constitutes only 
the second investigation of aging scaling at an active-to-absorbing phase
transition, and hence provides a crucial test of universality for 
non-equilibrium critical phenomena.

We also emphasize that universal aging scaling sets in considerably earlier 
during the system's time evolution than asymptotic quasi-stationary power laws
emerge, and in addition often yields more accurate exponent estimates 
\cite{gduct2012}.  
In comparison with detecting critical slowing-down, this critical aging effect 
might thus serve as a preferable and more reliable early warning signal for 
impending population collapse. 

\ack{This research is supported by the U.S. Department of Energy, Office of 
Basic Energy Sciences, Division of Materials Sciences and Engineering under 
Award DE-FG02-09ER46613.}

\end{document}